\documentclass{aa}
\usepackage{graphicx}
\usepackage{natbib}

\bibpunct{(}{)}{;}{a}{}{,}

\newcommand{\GILDAS}{\texttt{GILDAS}}

\newcommand{\IRAMthm}{\textrm{IRAM-30m}}

\newcommand{\CSO}{\textrm{\CSO}}

\newcommand{\eg} {{\em e.g.}}

\newcommand{\NHHD}  {\mbox{NH$_2$D}}         
\newcommand{\fNHHD}  {\mbox{$^{15}$NH$_2$D}}         

\newcommand{\Jone}{\mbox{$J$=1--0}}

\newcommand{\emm}[1]{\ensuremath{#1}}   
\newcommand{\emr}[1]{\emm{\mathrm{#1}}} 
\newcommand{\unit}[1]{\emm{\, \emr{#1}}}

\newcommand{\kms}   {\unit{km\,s^{-1}}}

\newcommand{\Tmb}{\emm{T_\emr{mb}}}
\newcommand{\Beff}{\emm{B_\emr{eff}}}
\newcommand{\Feff}{\emm{F_\emr{eff}}}

\newcommand{\TabObs}{%
  \begin{table}
    \caption{Source list }
    \begin{center}
      {\tiny
        \begin{tabular}{lcccl}
          \hline \hline
Source  & RA & Dec & V$_{LSR}$ & n(H$_2$)$^a$ \\
         & (2000) & (2000) & (\kms ) & (cm $^{-3}$)\\
\hline
Barnard 1b  & 03:33:20.9 & 31:07:34  & 6.8 & $3 \times 10^6$ \\
NGC1333-IRAS4A  & 03:29:10.5  & 31:13:31 & 7.2 & $2 \times 10^6$ \\
NGC1333-DCO$^+$   & 03:29:12.3  & 31:13:25 & 7.2 &  $1 \times 10^6$\\
L1544 & 05:04:16.6  &   25:10:48 & 7.4 &  $2 \times 10^6$\\ 
L134N(S)  & 15:54:08.6 & $-$02:52:10 & 2.4 & $2 \times 10^6$ \\
L1689N  & 16:32:29.5 & $-$24:28:53 & 3.8 & $2 \times 10^6$ \\
\hline
        \end{tabular}}
    \end{center}
$^a$ From Caselli et al. \cite{caselli08}
\label{tab:sources}
  \end{table}}

\newcommand{\TabTrans}{%
  \begin{table}
    \caption{Einstein coefficients, upper level energies and critical densities 
      for the range of temperatures considered in this work}
    \begin{center}
      \begin{tabular}{lccrrc} 
        \hline \hline
        Molecule  & Transition &   Frequency &    $A_{ij}$        
       &  $E_\emr{up}$  &    $n_\emr{crit}$     \\
            &      &    (GHz)        &      (s$^{-1}$)         &      (K)   
    &        (cm$^{-3}$)       \\
        \hline
o-\NHHD & $1_{1,1} - 1_{0,1}$ & 85926.2780 &  7.82e-6  &   20.68  & 4.2 10$^6$\\
o-\fNHHD & $1_{1,1} - 1_{0,1}$ & 86420.1959 & 7.96e-6 & 20.63 & 4.2 10$^6$ \\
p-\fNHHD & $1_{1,1} - 1_{0,1}$ & 109284.9021 & 1.61e-5  & 21.18 &8.8 10$^6$\\
        \hline
      \end{tabular}
     \label{tab:n_cr}
    \end{center} 
  \end{table}}

\newcommand{\TabResult}{%
  \begin{table*}
    \caption{Line intensities and molecular column densities}
    \begin{center}
      \begin{tabular}{l|c@{\hspace{1.5mm}}c@{\hspace{1.5mm}}c@{\hspace{3mm}}c@{\hspace{1.5mm}}c|c@{\hspace{1.5mm}}c@{\hspace{1.5mm}}cc@{\hspace{1mm}}|c}
        \hline \hline
 & \multicolumn{5}{c|}{o-\NHHD}   & \multicolumn{4}{c|}{o-\fNHHD}  & \\ 
 Source      &  $T_{mb} \pm \sigma$$^a$  & $\delta V$  &$\tau$ & $T_{ex}$ &  $N^b$  &  $T_{mb} \pm \sigma$$^a$  &  I & $\delta V$ &  
 $N^b$ &  ${[\NHHD]} \over {[\fNHHD]}$ \\
      & K  & kms$^{-1}$ & & K&  10$^{14}$ cm$^{-2}$ & mK  & mKkms$^{-1}$ &
kms$^{-1}$ &  10$^{11}$ cm$^{-2}$ &  \\ 
\hline
Barnard1b & $2.5 \pm 0.047$  & 0.79 & $5.24 \pm 0.14$ &$6.0 \pm 0.5$ & $4.7 \pm 0.5$  & $42 \pm 9$ 
  & $30 \pm 4$ & 0.67 & $10 \pm 2.7$ & $470^{+170}_{-100}$ \\ 
N1333-IRAS4A & $1.0 \pm 0.018$  & 1.38 & $1.39 \pm 0.10$ & $5.0 \pm 0.5$ & $2.7 \pm 0.6$ & $\pm 10$  & $< 30 $ & ... & $< 10 $ & $> 270 $  \\
N1333-DCO$^+$ & $1.3 \pm 0.015$  & 1.15 & $1.71 \pm 0.05$ & $ 5.3 \pm 0.5$ &
$2.4 \pm 0.4$ & $26 \pm 8$  & $14 \pm 3$ & 0.52 & $6.7 \pm 2.5 $  &  $360^{+260}_{-110}$ \\
LDN1544 & $2.3 \pm 0.016$  & 0.47 & $7.05 \pm 0.05 $ & $5.3 \pm 0.5$ & $4.1 \pm 0.5$ & $\pm 7$
   & $< 10 $ & ... & $< 5.2 $ & $> 700$ \\ 
L134N(S) & $2.2 \pm 0.033$  & 0.42 & $4.75 \pm 0.10$ & $5.5 \pm 0.5$ & $2.4 \pm 0.4 $ & $24 \pm 7 $
  & $10 \pm 2$ & 0.40 & $4.5 \pm 2$ & $530^{+570}_{-180}$ \\ 
L1689N & $5.3 \pm 0.030$  & 0.53 & $6.98 \pm 0.02$ & $8.5 \pm 0.5$ & $3.4 \pm 0.5$ & 
$65 \pm 17$  & $26 \pm 6$ & 0.37 & $4.2 \pm 1.5$ & $ 810^{+600}_{-250}$  \\
\hline
\end{tabular}
 \end{center}
$^a$ $\sigma$ is the rms computed for the original spectral resolution of 40 kHz = 0.136 \kms.

$^b$ computed at LTE with the same T$_{ex}$  for o-\NHHD \ and o-\fNHHD . 
T$_{ex}$ is derived from the HFS fit of the o-\NHHD \ profile.
     \label{tab:result}
  \end{table*}}

\newcommand{\Figspectra}{%
\begin{figure}
  \centering %
  \includegraphics[height=0.85\hsize{},angle=0]{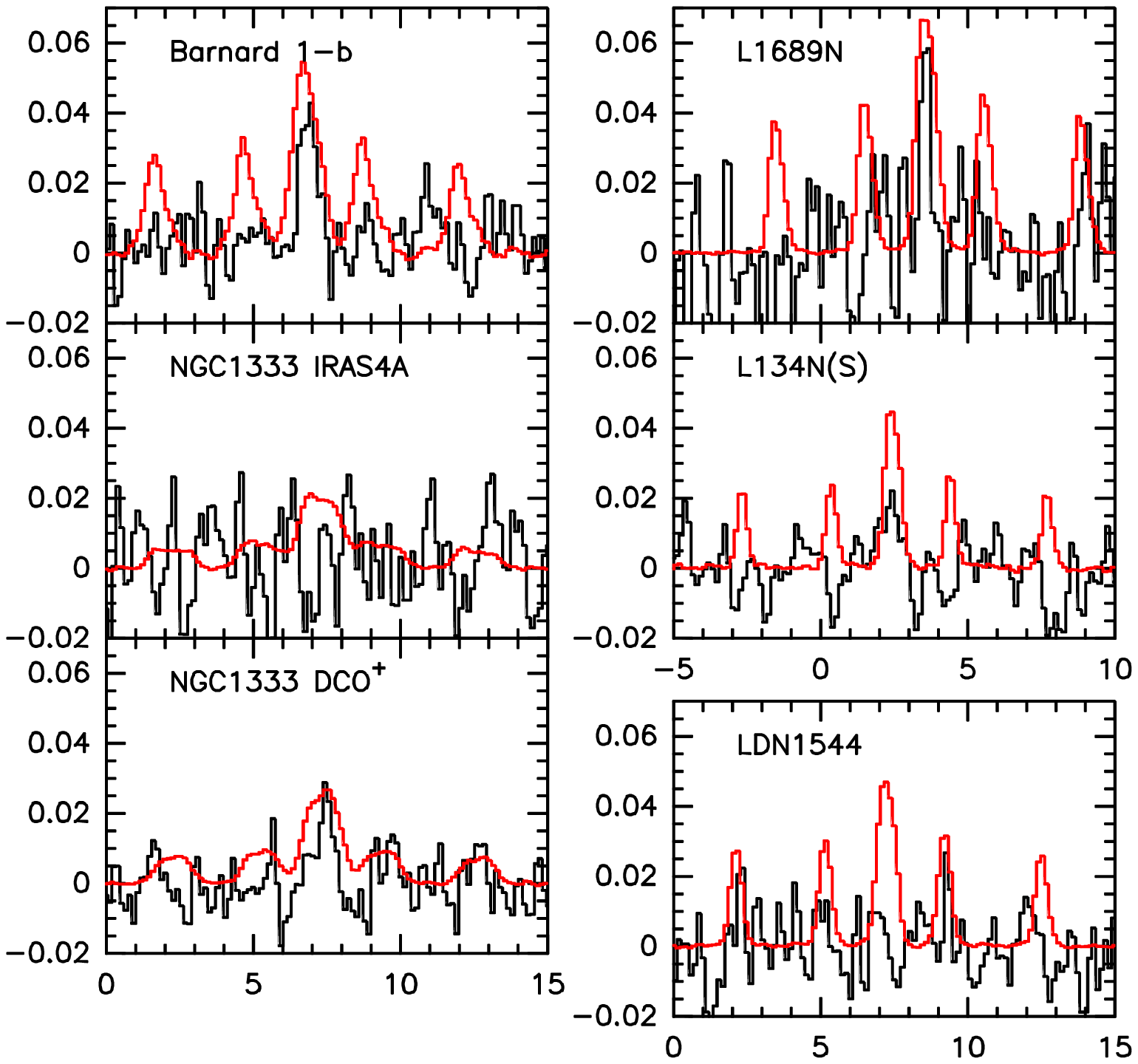}
   \caption{Spectra of the $1_{1,1}-1_{0,1}$ lines of o-\NHHD \ 
(grey/red line) 
and o-\fNHHD (black line) . The vertical scale is
\Tmb in K, the horizontal scale is V$_{LSR}$ in \kms . The o-\NHHD \ spectra have been multiplied by 0.02 except for the
L1689N spectra that has been scaled by 0.0125.  }
  \label{fig:spectra}
\end{figure}}

\newcommand{\FigModel}{%
\begin{figure}
  \centering %
  \includegraphics[height=\hsize{},angle=270]{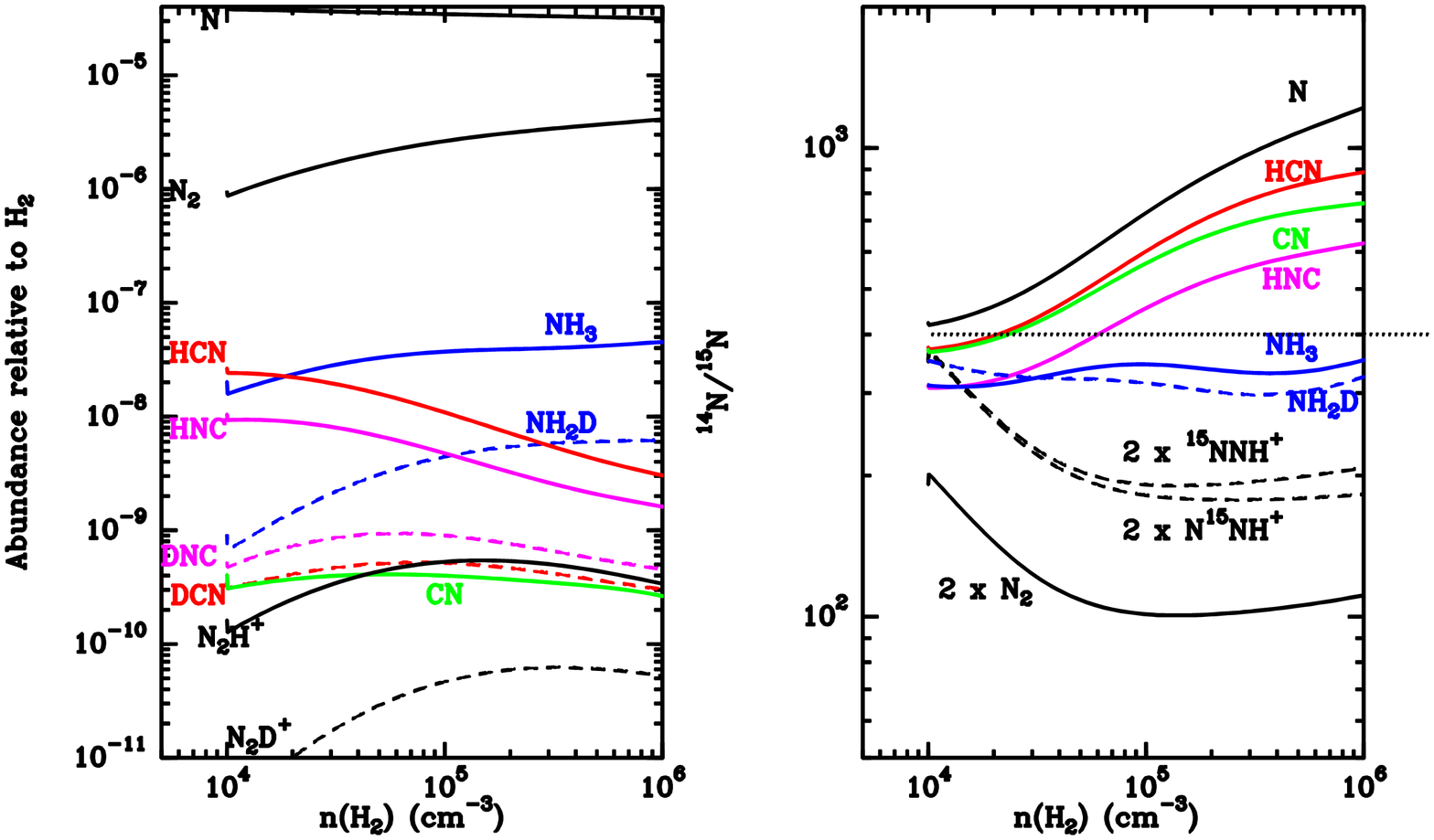}
   \caption{Prediction of the gas phase abundances relative to H$_2$ (left) and
$^{14}$N/$^{15}$N abundance ratio (right) for the main nitrogen species.
The models assumes a constant
temperature of 10K, and increasing depletions with the gas density,  to mimic
freezing out. The elemental abundance ratio $^{14}$N/$^{15}$N is set to 400.
 }
  \label{fig:model}
\end{figure}}

\begin{document}
\title{Detection of $^{15}$NH$_2$D in dense cores: A new tool for
measuring the $^{14}$N/$^{15}$N ratio in the cold ISM. \thanks{Based on
    observations obtained with the IRAM 
    30~m telescope. IRAM is supported by INSU/CNRS (France), MPG (Germany),
    and IGN (Spain).}}

\author{M. Gerin \inst{1}%
\and N. Marcelino \inst{2}
\and N. Biver \inst{3}
\and E. Roueff \inst{4}%
\and L. H. Coudert \inst{5}%
\and M. Elkeurti \inst{5}
\and D.C. Lis \inst{6}
\and D. Bockel\'ee-Morvan \inst{3}}

\offprints{\email{maryvonne.gerin@ens.fr}}

\institute{%
  LERMA, UMR 8112, CNRS, Observatoire de Paris and Ecole Normale
  Sup\'erieure, 24 Rue Lhomond, 75231 Paris cedex 05, France.
  \email{maryvonne.gerin@ens.fr} %
  \and
Laboratorio de Astrof\'{\i}sica Molecular, CAB-CSIC/INTA,
Ctra de Torrej\'on a Ajalvir km 4, 28850 Torrej\'on de Ardoz, Madrid, Spain. \email{nuria@damir.iem.csic.es}
\and
LESIA, UMR8109, CNRS and Observatoire de Paris, 5, place J. Janssen, 92195 Meudon
Cedex. France. \email{nicolas.biver@obspm.fr,dominique.bockelee@obspm.fr}
\and 
LUTh, Observatoire de Paris and UMR8102 CNRS, 5 place J. Janssen, 92195 Meudon 
Cedex. France. \email{evelyne.roueff@obspm.fr}
\and
LISA,  UMR 7583 CNRS and Universit\'e Paris 12, 61 Avenue du G\'en\'eral 
de Gaulle, 94010 Cr\'eteil Cedex, France. \email{coudert@lisa.univ-paris12.fr} %
\and
California Institute of Technology, MC 320-47, Pasadena, CA 91125, USA. \email{dcl@caltech.edu}
}

\date{Received / accepted }

\abstract
{Ammonia is one of the best tracers of cold dense cores. It is also a
  minor constituent of interstellar ices and, as such, one of the important
nitrogen reservoirs in the protosolar nebula, together with  the gas phase
nitrogen, in the form of 
N$_2$ and N.   An important diagnostic of the various nitrogen 
sources and reservoirs of
  nitrogen in the Solar System is the $^{14}$N/$^{15}$N isotopic ratio.
While good data exist for the Solar System, corresponding measurements in
the interstellar medium are scarce and of low quality.}
{Following the successful detection of the singly, doubly, and triply
  deuterated isotopologues of ammonia, we have searched for $^{15}$NH$_2$D in
dense cores, as a new tool for investigating the $^{14}$N/$^{15}$N ratio in
dense molecular gas. }
{With the IRAM-30m telescope, we have obtained deep integrations 
of the ortho $^{15}$NH$_2$D ($1_{1,1}-1_{0,1}$) line at 86.4 GHz, 
simultaneously with the corresponding ortho NH$_2$D line at 85.9 GHz.}
{The ortho $^{15}$NH$_2$D ($1_{1,0}-1_{0,1}$) is detected in Barnard-1b,
 NGC1333-DCO$^+$, and L1689N, while we obtained upper limits towards LDN1544
 and NGC1333-IRAS4A, and a tentative detection towards L134N(S). The
 para line at 109~GHz remains undetected at the  rms noise level achieved. The
 $^{14}$N/$^{15}$N abundance ratio in \fNHHD\ ranges between 350 and
 850, similar to {\bf the protosolar value} of $\sim 424$, and 
likely higher than the terrestrial ratio of $\sim 270$.}
{}


\authorrunning{Gerin et al.}
\titlerunning{Detection of interstellar $^{15}$NH$_2$D}
\maketitle{}

\Figspectra{} %
\TabObs{} %

\section{Introduction}

Nitrogen chemistry is particularly interesting for
understanding the connection between the ISM and the formation of the solar nebula,
because it is thought that the primitive atmospheres were nitrogen
rich, as Titan remains today. Furthermore, the isotopic
$^{15}$N/$^{14}$N ratio has been measured in a variety of Solar System
bodies, from the giant planets to the rocky planets, comets, and
meteorites. The observed differences in nitrogen fractionation are
used to understand how these bodies formed within the protosolar nebula.
The combination of nitrogen and hydrogen (D/H) isotopic ratios has
been demonstrated to be a very effective way of understanding how the
ice mantles were enriched in deuterium and nitrogen. Al\'eon and Robert 
\cite{aleon}
have concluded that a fast condensation of
the organic matter, enriched in $^{15}$N and deuterium, is needed in
order to keep a significant fractionation in the solid material of 
the primitive Solar System. They
also have evaluated the exothermicity of the fractionation reactions
for nitrogen to be $43 \pm 10$~K. The D fractionation has not been 
inherited from the native prestellar core, but most likely occurred in 
the protosolar nebula \citet{remusat,gourier}, 
yet the same physical and chemical
processes are thought to operate in the prestellar cores and in the 
coldest regions of circumstellar disks. 
In the ISM,  recent observations show that,
contrary to CO, nitrogen does not deplete from the gas phase in dense
cores, except when the density rises significantly above 10$^6$
cm$^{-3}$. Nitrogen species can therefore be very significantly
deuterated, with D/H fractionation of several tenths for N$_2$D$^+$ 
\cite{daniel07,pagani}  and \NHHD\ \citep{crapsi07}.
Multiply deuterated ammonia in particular  can be very abundant 
\citep{gerin06,lis02a,lis06,roueff05}.
Nitrogen molecules will therefore be
significant molecular reservoirs of deuterium. It is interesting 
to study whether
they could also be enriched in $^{15}$N, and whether signatures
from an enrichment at an
early evolutionary stage can be identified in primitive matter.

High $^{15}$N enhancements are measured both in HCN  and
CN cometary gases \cite{bockelee:08,schulz08}, {\bf and in primitive carbonaceous
meteorites. } 
High $^{15}$N enhancements may have
been present in the ammonia ices of the natal presolar cloud
according to the fractionation mechanism proposed 
by Rodgers \& Charnley \cite{rodgers:08a,rodgers:08b,rodgers:04} and
Charnley \& Rodgers \cite{charnley02}.
Nitrogen fractionation is not expected to be as efficient as deuterium
fractionation in dense cores, yet significant departures from the
elemental $^{14}$N/$^{15}$N ratio may occur in some molecules.
As first shown by Terzievia \& Herbst (2000), and developed 
by Charnley \& Rodgers \cite{charnley02} and Rodgers \& Charnley 
\cite{rodgers:08a}, nitrogen fractionation in the gas phase 
may operate through ion-molecule reactions
involving atomic or ionized nitrogen 
. 
Rodgers and Charnley \cite{rodgers:08b} have subsequently studied 
the possible role of 
neutral-neutral reactions involving $^{15}$N and CN. 
Little observational interstellar data
are available.
We have therefore started a survey of the main nitrogen-bearing 
interstellar species in 5 dense cores and a class 0 source (Table~\ref{tab:sources}). 
This paper reports the detection
of o-\fNHHD \ as the first result of this survey.

\section{Observations}
\label{sec:obs}
\TabTrans{}

The microwave and far infrared spectra of \fNHHD\ and  $^{15}$NHD$_2$ 
have been recently investigated by Elkeurti et al. \citep{elkeurti:08} 
and used to produce the corresponding line lists as supplementary data\footnote{Available at
\texttt{http://library.osu.edu/sites/msa/suppmat/ v251.i1-2.pp90-101/mmc1.txt}}, while accurate line lists and partition
functions for the $^{14}$N isotopologues of the NH$_3$ family can
be found in Coudert \& Roueff \cite{coudert}. These species are also independently  listed in the Cologne
Database for Molecular Spectroscopy (CDMS, M\"uller et
 al. 2001, 2005),
with small differences in the line frequencies due to different handling of the hamiltonians.

We have chosen to search for the $1_{1,1}-1_{0,1}$ line of ortho
\fNHHD\, since
the corresponding \NHHD\ line is very strong and both the sky transmission and
telescope performances are  excellent {\bf at 86~GHz}. The frequency shift 
introduced by the $^{15}$N substitution is small enough that the 
two isotopologues can be observed with the same receiver tuning. The line
frequencies (Elkeurti et al. 2008), Einstein A 
coefficients, upper energy 
levels and critical
densities are listed in Table~\ref{tab:n_cr}. We have used the theoretical
estimates of the critical densities from the reduced mass ratio scaling  
of Machin \& Roueff \cite{machin} for the \NHHD --He values at 10~K, the
temperature appropriate for the cold cores we have observed. However these 
values are 
probably too large when molecular hydrogen is involved, as found in recent calculations
 of the NH$_3$-H$_2$ system by Valiron et al. (private communication).

The observations have been performed with the IRAM-30m telescope, during three
observing sessions in December 2007, March 2008 and September 2008.  We used
the A100 and B100 receivers in parallel, tuned to 86.2 GHz in order to 
detect o-\NHHD\ and
o-\fNHHD\ with the same detector setting. The weather conditions were average, 
with $5 - 10$~mm of water vapor (PWV) .
 {\bf The  \NHHD\ and \fNHHD\ lines were observed simultaneously,
with  the \Jone \ lines of
H$^{15}$NC and  H$^{13}$CN at 86.055~GHz and  86.338~GHz. 
We used the  VESPA correlator,}
tuned to a spectral resolution of 40~kHz, and spectral bandpass of
40 MHz for each line.
The data were taken using the wobbling secondary reflector, with a
beam separation of $240 "$. Telescope pointing was checked on nearby
planets and bright radio quasars and was found accurate to $\sim 3 ''$. 
Due to rather poor weather conditions during the September run (high PWV
 and cloudy sky), the pointing accuracy was degraded to $\sim 5 ''$.
Additional observations of the p-\fNHHD\ line at 109.3 GHz were obtained
in March 2008. We only searched for this line towards Barnard-1b and
detected no signal down to a rms noise level of 18~mK  with 0.2~\kms\ 
velocity resolution. For L134N(S), we combined the data with observations performed in April 2005, as part of the dark cloud line survey project \cite{marcelino}. 
The weather conditions were excellent (1 -- 2 mm PWV) and the
observations performed in  the frequency switching mode.

The data processing was done with the \GILDAS\footnote{See
  \texttt{http://www.iram.fr/IRAMFR/GILDAS}} software (\eg{} Pety et al. 2005).
We used the dec08b version of this software, which allows to correct for a minor bug in the frequency calibration during the observations.
The \IRAMthm{} data 
are presented in main beam temperatures $T_{mb}$,
 using the forward and main beam
efficiencies \Feff{} and \Beff{} appropriate for 86~GHz, \Feff =0.95 and
\Beff =0.78.  The {\bf uncertainty in flux calibration} is $\sim 10\%$, as
checked by the variation of the intensity of the strong o-\NHHD\ and 
H$^{13}$CN lines in the spectrum. Linear baselines were
subtracted.

Because the nuclear spin of $^{15}$N is 1/2, the \fNHHD\ 
lines are split
into fewer hyperfine components than \NHHD\, which makes their detection more
favorable. The hyperfine structure of \fNHHD \  
is driven by the quadrupole  
moment of the deuterium nucleus, which is much smaller than the  
corresponding value of $^{14}$N.  We have checked, by using the  
nuclear quadrupole constants provided in Garvey et al. \citet{garvey}, that  
the resulting hyperfine splitting is less than 50~kHz. We can thus  
safely assume that the  spectrum reduces to a single  
component
. As shown in Figure~\ref{fig:spectra}, the \fNHHD\ line is 
clearly detected towards Barnard-1b,
 and L1689N, while we obtained upper limits towards LDN1544 and
NGC1333-IRAS4A and {\bf tentative detections towards NGC1333-DCO$^+ $ and L134N(S)}. 
The ratio of peak antenna temperatures {\bf of the  \NHHD\ and \fNHHD\ lines  is 50 - 100,
and  the velocity  agreement is excellent.}
Using the JPL and
CDMS spectroscopy data bases, we have checked that no line of
 known interstellar molecules  are expected within $\pm 300$~kHz from
the \fNHHD \ line frequency. The identification of the detected feature 
is therefore secure.   

The line parameters were estimated by fitting Gaussian profiles to
the detected o-\fNHHD\ lines. For o-\NHHD\, we used the HFS routine 
implemented in 
CLASS, which allows to take into account the hyperfine components self-consistently. The opacity of
the ortho \NHHD \ line is moderate in all sources, with a total
opacity for all lines ranging from $\sim 1$  to $\sim 5$ 
(Table~\ref{tab:result}). 

\section{Results}
\label{sec:result}

\subsection{\NHHD \  and \fNHHD}
\TabResult{}

Results of the fits and derived molecular column densities are listed
in Table \ref{tab:result}. As we are mostly interested in the
ratio of column densities, we computed them under
the simple assumption of a single excitation temperature. We used
the excitation temperature derived from the \NHHD \ fit for both
isotopic species. 
The o-\NHHD\ column densities are in good agreement with previously published
results for the same sources \cite{roueff05}.
The [\NHHD]/[\fNHHD] abundance ratio range from 360 to 810, with the largest
value for L1689N.  This last source is an interaction region between a molecular outflow and a dense core, and as such may have peculiar properties 
\cite{lis02b}.   Given the  error bars, the measured 
[\NHHD]/[\fNHHD] ratio is comparable to the {\bf $^{14}$N/$^{15}$N 
 protosolar
ratio, as measured in Jupiter (450; Fouchet et al. 2004)
and in  osbornite-bearing calcium-aluminium-rich
  inclusions from meteorites
(424; Meibom et al. 2007),  and likely larger than the terrestrial
abundance ratio (270)}.  Although the uncertainty on the [\NHHD]/[\fNHHD] ratio 
remains large, the cold prestellar cores L1689N and LDN1544  seem to have
a larger ratio than  Barnard-1b and NGC1333-DCO$^+$.

\subsection{$^{15}$N fractionation}
Nitrogen fractionation  involves two main mechanisms in the gas phase: 
isotopic dependent photodissociation of molecular N$_2$, principally at 
work in the atmosphere of Titan (Liang et al. 2007) and 
possible ion-molecule fractionation reactions occurring at low temperatures 
in cold dense cores as first measured by Adams \& Smith 
(1981) and calculated by Terzieva and Herbst (2000).  
In this latter case, involved endothermicities 
values range between a few K up to 36 K 
for exchange reactions involving $ ^{15}$N, $ ^{15}$N$^+$, and $^{15}$NN.
{\bf Selective photodissociation of N$_2$ and $^{14}$N$^{15}$N takes place at  
wavelengths between 80 and 100~nm, a range where cold dense cores
 are completely opaque. Then, this mechanism does not  
work in the present context.}
Charley \& Rodgers \cite{charnley02} and Rodgers \& Charnley \cite{rodgers:08a} have investigated the nitrogen fractionation in their
time dependent, coupled gas/solid chemical models. They conclude
that $^{15}$N-rich ammonia and deuterated ammonia
can be frozen onto the ice mantles,  provided all nitrogen is not 
converted into N$_2$. The gas phase becomes enriched at early times,
before the complete freezing of the gas phase molecules.

Additional fractionation reactions may be introduced such as those 
involving $ ^{15}$N$^+$ with CN and NH$_3$ 
and some neutral-neutral reactions between $^{15}$N and 
CN (Rodgers \& Charnley 2008b).
However, none of these reactions has been studied in the laboratory and
these schemes remain highly hypothetical.
We have developed a gas  phase chemical code, 
including ion-molecule  fractionation reactions for carbon and nitrogen
\cite{langer:92,langer:84,terzieva00}
as well as a complete deuterium chemistry (Roueff et al. 2005). 
We have explicitly introduced D and $^{13}$C on the one hand and D and $^{15}$N on the other hand  for NH$_n$, HCN and HNC molecules, in order to directly 
compare the model results with the observations.
 The chemical network involves  302
chemical species and 5270 reactions. The maximum number of carbon atoms 
in a molecule  has been limited to 3.
We have introduced the additional
 reaction channels  arising from the inclusion of isotopic species. 
We have also preserved
 functional groups in dissociative recombination reactions such as :
 \begin{equation}
\rm  \mbox{HCND}^+ +e \rightarrow \mbox{HCN} + D 
 \end{equation}

 \begin{equation}
\rm  \mbox{HCND}^+ +e  \rightarrow  \mbox{DNC} + H 
 \end{equation}
 
Note that the branching ratios of the 
dissociative recombination of N$_2$H$^+$ have been measured again by 
Molek et al. (2007) with the result that the channel towards N$_2$ occurs 
with a probability of at least 90\% 
\FigModel{}

A  calculation is shown in Figure~\ref{fig:model} for typical dense core
parameters, and assuming a $^{14}$N/$^{15}$N abundance ratio of 400, and
an ionization rate of $\zeta = 2 \times 10^{-17}$ s$^{-1}$.
 The model predicts that the  $^{15}$N enrichment of ammonia
is moderate in the gas phase, while a stronger enrichment is predicted for
N$_2$H$^+$, and depletion for HCN and CN. Recent models by Rodgers and
Charnley (2008a) obtain similar results for the gas phase 
abundances, the
$^{15}$N enrichment of ammonia being more efficient in the solid phase.

\section{Conclusions}
\label{sec:conclusion}
{\bf We report the detection of heavy deuterated ammonia, \fNHHD , in 
three cold dense cores. The abundance ratio [\NHHD]/[\fNHHD] is compatible
with the $^{14}$N/$^{15}$N protosolar value, and seems larger 
than the terrestrial value despite the remaining  measurement uncertainties.}
While further observations are
needed to {improve the accuracy and} test our chemical models, 
ammonia and deuterated ammonia seem
to be good probes of the $^{14}$N/$^{15}$N ratio. Deuterated ammonia
is particularly interesting as it probes the coldest and densest
regions of prestellar cores which are the reservoirs for the
future formation of young stars and their associated protoplanetary disks.

\begin{acknowledgements}
  We thank the IRAM director for assigning additional time for this 
program, which helped us to
  confirm the \fNHHD\ detection and the 30m staff for their support during the
  observations. We thank the referee, E. Bergin, for his insightful comments.
We acknowledge financial support from the
CNRS interdisciplinary program ``Origines des Plan\`etes et de la Vie'', 
and from the INSU/CNRS program PCMI. 
NM is supported by Spanish MICINN
through grants AYA2006-14876, by DGU of the Madrid
community government under IV-PRICIT project S-0505/ESP-0237
(ASTROCAM), and by Molecular Universe FP6 MCRTN."
DCL is supported by the
NSF, grant AST-0540882 to the Caltech Submillimeter Observatory.
\end{acknowledgements}

\end{document}